\documentclass[preprint,review,12pt]{elsarticle}

\usepackage{graphicx}
\usepackage{natbib}
\usepackage{epstopdf}
\usepackage{graphicx, subfigure}
\usepackage{subeqnarray}
\usepackage{amsmath}
\usepackage{amssymb}
 \usepackage{amsthm}

\begin{document}

\begin{frontmatter}



\title{Oscillatory instability of fully 3D flow in a cubic diagonally lid-driven cavity.}


\author{Yu. Feldman}
\ead{yurifeld@caltech.edu}
\tnotetext[label1]{telephone 1-626-863-6802, fax 1-626-568-2719}
\address{California Institute of Technology, Pasadena, CA 91125, USA}

\begin{abstract}
A transition to unsteadiness of a flow inside a cubic diagonally lid-driven cavity with no-slip boundaries is numerically investigated by a series of direct numerical simulations (DNS) performed on $100^3$ and $200^3$ stretched grids. It is found that the observed oscillatory instability is setting in via a subcritical symmetry-breaking Hopf bifurcation. The instability evolves on two vortices in a coupled manner. Critical values of Reynolds number $Re_{cr}=2320$ and non-dimensional angular oscillating frequency $\omega_{cr}=0.249$ for transition from steady to oscillatory flow are accurately estimated. Characteristic patterns of the 3D oscillatory flow are presented.
\end{abstract}


\end{frontmatter}


\section{Introduction}
Lid-driven cavity flow has been the subject of intensive theoretical and experimental research for many decades. Such tremendous scientific interest is due to the overwhelming importance of this kind of flow to the basic study of fluid dynamics \cite{shankar2000anrev}. Started by the early theoretical works of \cite{batchelor1956jfm} and \cite{moffatt1963jfm}, and followed by the numerical studies of \cite {kawaguti1961jpsj} and \cite{simuni1965jamtp} the state of the art lid-driven flow research represents the whole diversity of fluid transport phenomena. It includes longitudinal vortices, corner eddies, non-uniqueness, transition to unsteadiness and turbulence \cite{shankar2000anrev}.

\begin{figure}
\centering
    {
        \includegraphics[width=0.6\textwidth,clip=]{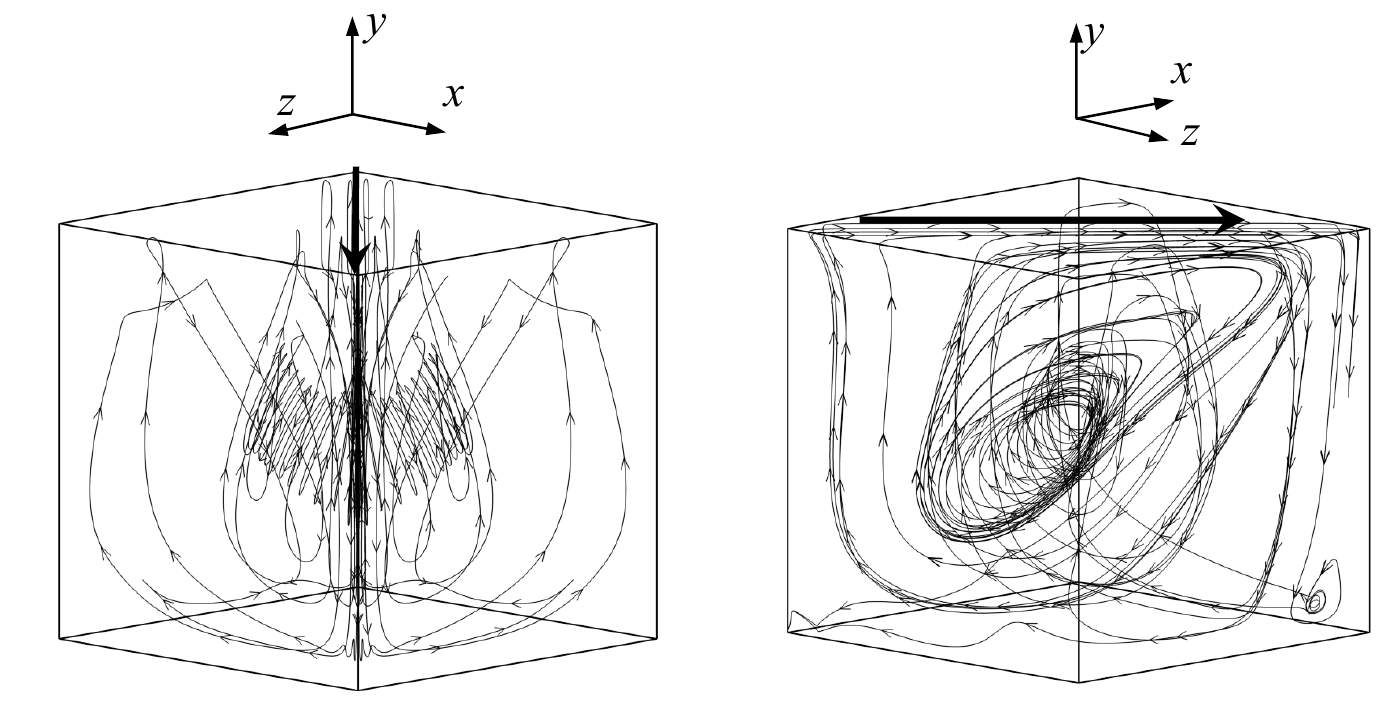}
    }
    \caption{Flow path lines in a cubic lid-driven cavity with lid moving at $45^\circ$ to the $x$ axis, $Re = 1000$ \cite{feldman2010computfluid}, lid moves as indicated by an arrow.}
\label{fig:intro}
\end{figure}

The ``classical'' lid-driven cavity flow, though comprising very popular benchmark for verification of numerical methods and validation of experimental methodologies, has limitations. Chief among these is that, although realized for 3D geometry, this flow still has 2D similarities. In fact, steady state lid-driven cavity flow in a cubic box is symmetric relatively to the cavity midplane (see e.g. \cite{feldman2010physfluid}) with prevailing 2D character in its vicinity.  The flow symmetry breaks for unsteady slightly supercritical regime \cite{feldman2010physfluid} which nevertheless does not result in a significant change in its 2D character (magnitude of spanwise, $z$ velocity component remains small relatively to the magnitudes of both $x$ and $y$ velocity components). The discussed limitation was remedied by formulation of an alternative benchmark problem simulating flow inside cubic diagonally lid-driven cavity \cite{povitsky2005}. Contrary to its ``classical'' analogue, this flow is a priori fully 3D (see Fig. \ref{fig:intro}). For steady state regime  the flow is symmetric relatively to diagonal plane and is characterized by the same velocity components in $x$ and $z$ directions \cite{feldman2010computfluid}.

Over the recent decade the steady state diagonally lid-driven flow has become a popular benchmark for verification of state of the art numerical methods \cite {dazhi2003,dhumi2011PhilTrans,freitas2011CompFluids,asinari2012jcp}. However, neither transition to unsteadiness  mechanism  nor characteristics of the slightly supercritical regime of this interesting flow have so far been addressed by anyone. The later is studied by linear stability analysis aimed at identification of critical conditions for the instability and classification of its 3D eigenmodes. The research in this field has been mainly motivated by a strong overestimation of the 2D flow stability limit compared to 3D configurations (see e.g. \cite{ramanan1994physfl} and \cite{theofilis2000AIAA}), rendering the 2D flow results unsuitable for quantitative comparison with the 3D experimental data. First accurate results for instability analysis for a 3D lid-driven cavity flow with spatially periodic spanwise boundary conditions were reported in \cite{ding1998JCP}. The results were confirmed and extended by the studies of \cite{albensoeder2001physfluid} and \cite{theofilis2000AIAA}, who independently determined a prevalence of stationary leading mode followed by three different traveling modes. The first global instability analysis (with all no-slip boundaries) for cubic lid-driven cavity flow was recently presented by \cite{gianettiatti} who addressed a 3D eigenvalue problem on a $128^3$ spectral collocated grid. The authors revealed a spatial structure of leading eigenmode existing at $Re=2000$. Independently, an oscillatory instability in this flow  setting in via a subcritical symmetry-breaking Hopf bifurcation at $R_{cr}=1914$ was reported in \cite{feldman2010physfluid}. To characterize the observed instability mode the authors presented  a spatial distribution of the velocity amplitudes, which where then favorably verified by the experimental results published in \cite{liberzon2011physfluid}.

This study extends global instability analysis to the case of a diagonally lid-driven cubic cavity (see Fig. \ref{fig:intro}). We are the first to report accurate zero-grid-size limit values of critical Reynolds number $Re_{cr}=2320$ and angular oscillating frequency $\omega_{cr}=0.249$ for steady-unsteady transition, obtained by applying Richardson extrapolation to the corresponding values calculated on $100^3$ and $200^3$ cartesian stretched grids. We report  that the transition to unsteadiness takes place via symmetry-breaking oscillatory subcritical Hopf bifurcation, and discuss the the instability type. In addition, spatial distributions of velocity and vorticity oscillating amplitudes, useful for verification of future linear stability results, are presented.
\section{Computational details and verification}
A cubic lid-driven cavity with side of length $L$ is considered. The cavity top lid is moving with a constant velocity $U$ at $45^\circ$ to the cube's vertical walls (see Fig. \ref{fig:intro}). All other boundaries of the cavity are stationary. The flow is governed by the incompressible continuity and momentum equations with no-slip boundary conditions applied on all the boundaries:
\begin{subeqnarray}
 \nabla\cdot \textbf{\textit{\textbf{u}}} =0,\\[3pt]
 \frac{\partial{\textbf{\textit{\textbf{u}}}}}{\partial{\textit{t}}}+(\textbf{\textit{\textbf{u}}}\cdotp\nabla){\textbf{\textit{\textbf{u}}}}=
-\nabla{p}+\frac{1}{Re}\nabla^2 \textbf{\textit{\textbf{u}}},
  \label{govern}
\end{subeqnarray}
where velocity vector $\textbf{u}(u,v,w)$, pressure $p$, time $t$ and all length scales are normalized by $U$, $\rho U^2$ ($\rho$ is the fluid density), $L/U$ and $L$ respectively. The Reynolds number is defined as $Re=UL/\nu$, where $\nu$ is the kinematic viscosity of the fluid. The governing equations (\ref{govern}) were solved with an icoFoam solver, which is part of an open source parallelized code openFoam \cite{weller1998}. The simulations were performed on a standard unix cluster and involved up to 512 cores running in parallel. SIMPLE algorithm (see e.g. \cite{ferzirger2002}) was used for pressure-velocity coupling and a conservative second order finite volume scheme was utilized for the spatial discretization. The time derivative in the momentum equations was approximated by the second order backward finite difference. A zero pressure gradient normal to all the walls was assumed when solving Poisson's equation.

Verification of the icoFoam solver was performed by comparison of the obtained results with the data previously published in \cite{feldman2010computfluid} for the same flow configuration. Table \ref{tab:verific} summarizes the results of both studies for the velocity and pressure fields monitored for $Re=1000$ along the cavity vertical centerline.
\begin{table}
\begin{center}
\begin{minipage}{12.7cm}
\begin{tabular}{@{}ccccccl@{}}
{y} & {$v_{x}, v_{z} \times 10^3$} & $ $ &{$v_{y} \times 10^3$} & $ $ &{$p \times 10^4$}& \multicolumn{1}{c@{}}{}\\[3pt]
$ $ & $Ref.$  & $Pres.$ &$Ref.$  & $Pres.$ &$Ref.$  & $Pres.$\\
$0.9766$ & $417.7$ &   $417.8$  & $5.378$ &  $5.458$ & $51.59$ & $51.42$\\
$0.9531$ & $226.6$ &   $226.5$  & $16.07$ &  $16.14$ & $46.67$ & $46.60$\\
$0.8516$ & $76.74$ &   $76.39$  & $30.36$ &  $30.36$ & $35.18$ & $35.00$\\
$0.7344$ & $62.50$ &   $62.00$  & $22.59$ &  $22.50$ & $21.85$ & $21.73$\\
$0.6172$ & $41.78$ &   $41.22$  & $5.790$ &  $5.561$ & $9.711$ & $9.664$\\
$0.5000$ & $-1.398$ &$-1.395$ & $-33.95$&  $-34.08$  & $0.000$ & $0.000$\\
$0.4531$ & $-31.54$ &$-31.33$ & $-64.70$ & $-64.35$  &$-2.517$& $-2.4525$\\
$0.2813$ & $-130.7$ &$-129.0$ & $-160.2$ & $-158.0$  & $28.24$ & $28.03$\\
$0.1719$ & $-134.7$ &$-133.6$ & $-137.9$ & $-135.9$  & $107.0$ & $106.0 $\\
$0.1016$ & $-143.1$ &$-142.5$ & $-86.78$ & $-85.41$  & $186.5$ & $184.5$\\
$0.0547$ & $-162.2$ &$-161.1$ & $-35.52$ & $-34.49$  & $232.9$ & $230.6$
\end{tabular}
\end{minipage}
\end{center}
\caption{Pressure and velocity values along the cavity centerline (0.5, y, 0.5), $Re=1000$ : comparison between the reference \cite{feldman2010computfluid} ($152^3$ grid) and the present ($100^3$ grid) results}
\label{tab:verific}
\end{table}
Deviation between the corresponding velocity and pressure values does not exceed 1\% favorable verifying the present results. Note also the same values (up to 6-th decimal digit) of $v_{x}$ and $v_{z}$ velocity components which indicate a reflection symmetry of the obtained steady state flow, in agreement with the previous studies of \cite{povitsky2005} and \cite{feldman2010computfluid}.
\section{Results}
\subsection{Transition to unsteadiness}
The transition to unsteadiness was investigated by simulating the subcritical flow over small increments of $Re$ number until at $Re>Re_{cr}$ the steady flow broke down superposed by a periodic flow motion. Given that this periodic secondary flow is dictated by small amplitude (linear) dynamics, the $Re_{cr}$ value can be regarded as a Hopf bifurcation point. Mathematically, this means that the spectrum of a linearized set of equations has only a single pair of complex eigenvalues $\sigma\pm i\omega$ whose real part $\sigma$ is crossing the axis of neutral stability \citep{drazin2004}. In that case the dynamics of slightly supercritical flow close to the bifurcation point is described by the Hopf theorem \citep{hassard1981}:
\begin{equation}
  v(t,Re)=v_{0}(Re_{cr})+\epsilon Real(\textbf{V} e^{i\omega t})+\textit{O}(\epsilon^2), \frac{\partial \sigma}{\partial Re}|_{Re_{cr}}\neq0.
  \label{Hopf}
\end{equation}
Here $v_{0}$ is the base (steady) flow at $Re=Re_{cr}$ and $\textbf{V}$ is the leading eigenvector corresponding to the leading eigenvalue $i\omega_{cr}$. If the observed Hopf bifurcation is supercritical, then a stable  continuously growing limit cycle exists in the very vicinity of the critical point and both the oscillation amplitude $\epsilon$ and the deviation of the oscillations frequency from its critical value $\omega-\omega_{cr}$, are proportional to $\sqrt{Re-Re_{cr}}$ (see e.g. \cite{hassard1981}). Then a sequence of $\epsilon$ and $\omega$ values acquired for two slightly supercritical flows would provide an acceptable approximation for $Re_{cr}$ and $\omega_{cr}$.

On the contrary, the subcritical bifurcation involves an unstable limit cycle which can not be directly reproduced by a time integration close to the bifurcation point. It is distinguished by an abrupt discontinuous increase of oscillation amplitude $\epsilon$ from zero ($Re<Re_{cr}$) to some finite value ($Re>Re_{cr}$). It can also have a hysteresis region which is characterized  by different values of critical Reynolds number, $Re_{cr1}>Re_{cr2}$, where $Re_{cr1}$ and $Re_{cr2}$ are related to stationary-oscillatory and oscillatory-stationary transitions respectively. Both of the above characteristics were observed in our numerical simulations. Moreover, applying Hopf theorem (\ref{Hopf}) to two consequent solutions located on a stable branch of the limit cycle resulted in a considerable (about 10\%) underestimation  of $Re_{cr}$ value compared to that tracked by small $Re$ increments. All of the above leads to the conclusion that the observed bifurcation is of subcritical type.

Adapting an approach recently applied by \citep{feldman2010physfluid}, estimation of the critical values was performed by analysis of a series, corresponding to subcritical flow regimes characterized by decaying oscillation amplitudes. It is based on the observation that, after a long enough time, the close to bifurcation point subcritical flow is dictated only by the most unstable mode while the flow oscillations $f(t)$ decay proportionally to {$e^{\sigma+i\omega t}$, $\sigma<0$  whereas the value of $\sigma$ is calculated by:
\begin{equation}
 \sigma=\frac{ln(f(t_{k}/f(t_{k-1})}{t_{k}-t_{k-1}}.
  \label{Subcrit}
\end{equation}
Here $t_{k}$ $(k=1,2,3,...)$ correspond to the instant times when the flow oscillations $f(t_{k})$ attain their local maxima. The values of $Re_{cr}$ and $\omega_{cr}$ are then calculated by extrapolation of $\sigma$ to the zero value. Time evaluations of $v_{x}$ component monitored for the values of $Re=2300$ and $Re=2325$ within the region with the largest oscillation amplitudes (control point $A(0.66, 0.42, 0.56)$) are shown in Fig.\ref{fig:Evaluation}.
\begin{figure}
\centering
    \subfigure[]
    {
        \includegraphics[width=0.4\textwidth,clip=]{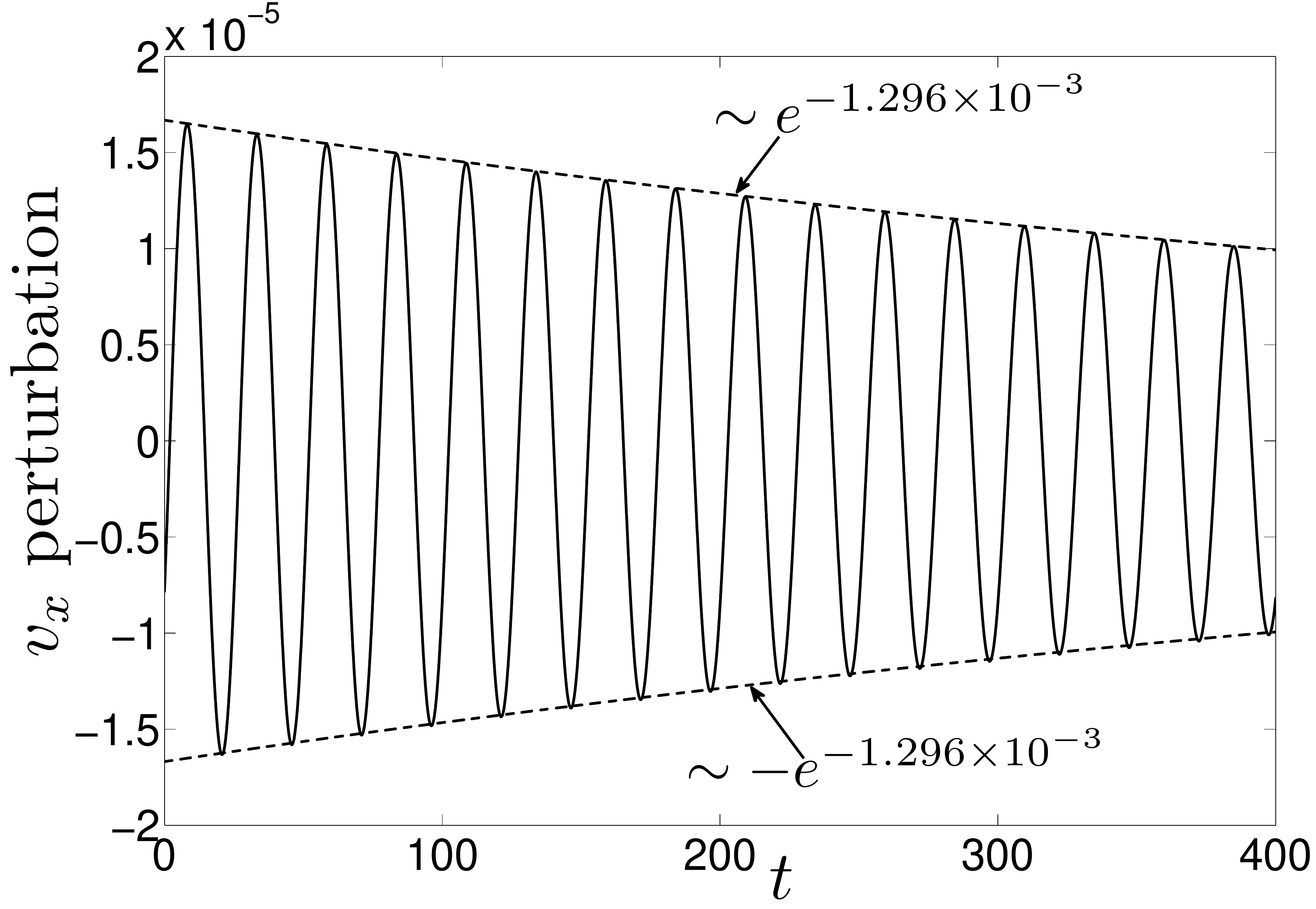}
    }
    \subfigure[]
    {
        \includegraphics[width=0.4\textwidth,clip=]{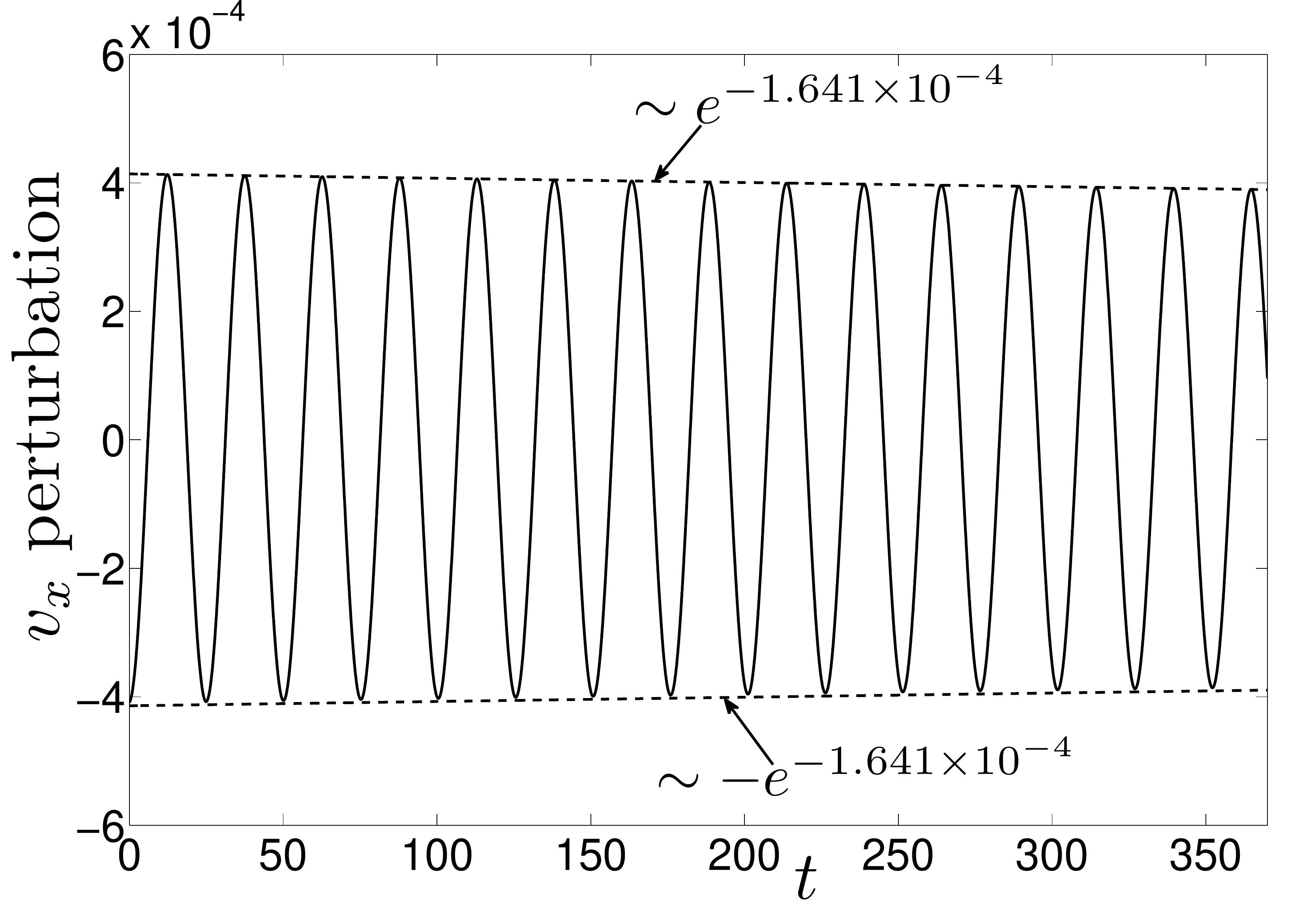}
    }
    \caption{Time evaluation of the $v_{x}$ velocity component monitored at a control point (0.66, 0.42, 0.56), $100^3$ grid: (a) $Re$=2300; (b) $Re$=2325.}
\label{fig:Evaluation}
\end{figure}

As expected, a lower decay ratio value corresponds to the higher $Re$ which was chosen to be very close to the bifurcation point. The same calculations were repeated for a refined grid containing $200^3$ finite volumes. The critical values calculated on the both grids are reported in Table \ref{tab:Critical}. For all cases, the precision of $\sigma$ and $\omega$ values was verified up to the third decimal digit. We also checked by Fourier analysis (not shown here) that the frequency spectrum of all the signals contain only a single value corresponding to a leading mode, while the disturbances introduced by other modes have already adequately decayed. Following the works of \citep{gelfgat2007Int_j,feldman2010physfluid} Richardson extrapolation was applied for the further improvement of the grid-dependent $Re_{cr}$ and $\omega_{cr}$ results to their zero-grid-size asymptotic values, yielding $Re_{cr}=2320$ and $\omega_{cr}=0.249$ values to three decimal places.
\begin{table}
\begin{center}
\begin{tabular}{@{}cccl@{}}
{Grid resolution} & {$Re_{cr}$} & {$\omega_{cr}$}
& \multicolumn{1}{c@{}}{}\\[3pt]
$100^3$ & $2329$&  $0.2495$ \\
$200^3$ & $2321$ & $0.2488$ \\
$Richardson \ extrapolation$ & $\textbf{2320}$ & $\textbf{0.249}$
\end{tabular}
\end{center}
\caption{Estimation of  $Re_{cr}$ and $\omega_{cr}$ values}
\label{tab:Critical}
\end{table}

\subsection{Slightly supercritical flow field}
A slightly supercritical flow field in the very vicinity of the subcritical Hopf bifurcation point, $Re=2335$ is considered. After a sufficiently long time, the spectrum of this flow consists of only a single frequency value (and its multipliers caused by non-linear effects) corresponding to the unstable mode, while all the disturbances initially introduced into the system are being damped to the machine zero \cite{feldman2010physfluid}. This is true for any velocity component monitored at any internal point of the confined volume. To get more insight into the characteristics of the observed slightly supercritical flow we look at the projections of $v_{x}$ and $v_{z}$ velocity components on the $X-Z$ plane diagonals:
\begin{subeqnarray}
 v_{dir}=(v_{x}+v_{z})cos(\pi/4),\\[3pt]
 v_{perp}=(v_{x}-v_{z})sin(\pi/4),
  \label{v_dir}
\end{subeqnarray}
where $v_{dir}$ is directed in the same direction with the cavity lid motion and $v_{perp}$ is perpendicular to $v_{dir}$. Fig. \ref{fig:evaluations} shows the time evaluations and the corresponding frequency spectra of the newly defined quantities along with the $v_{y}$ component monitored at three control points: a pair of symmetry reflection points A(0.66, 0.42, 0.56) and B(0.56 0.42,0.66) with respect to the cavity main diagonal plane and an independently chosen point C(0.1 0.14 0.1) located on the main diagonal plane.
\begin{figure}
\begin{center}
\includegraphics[width=0.9\textwidth,clip=]{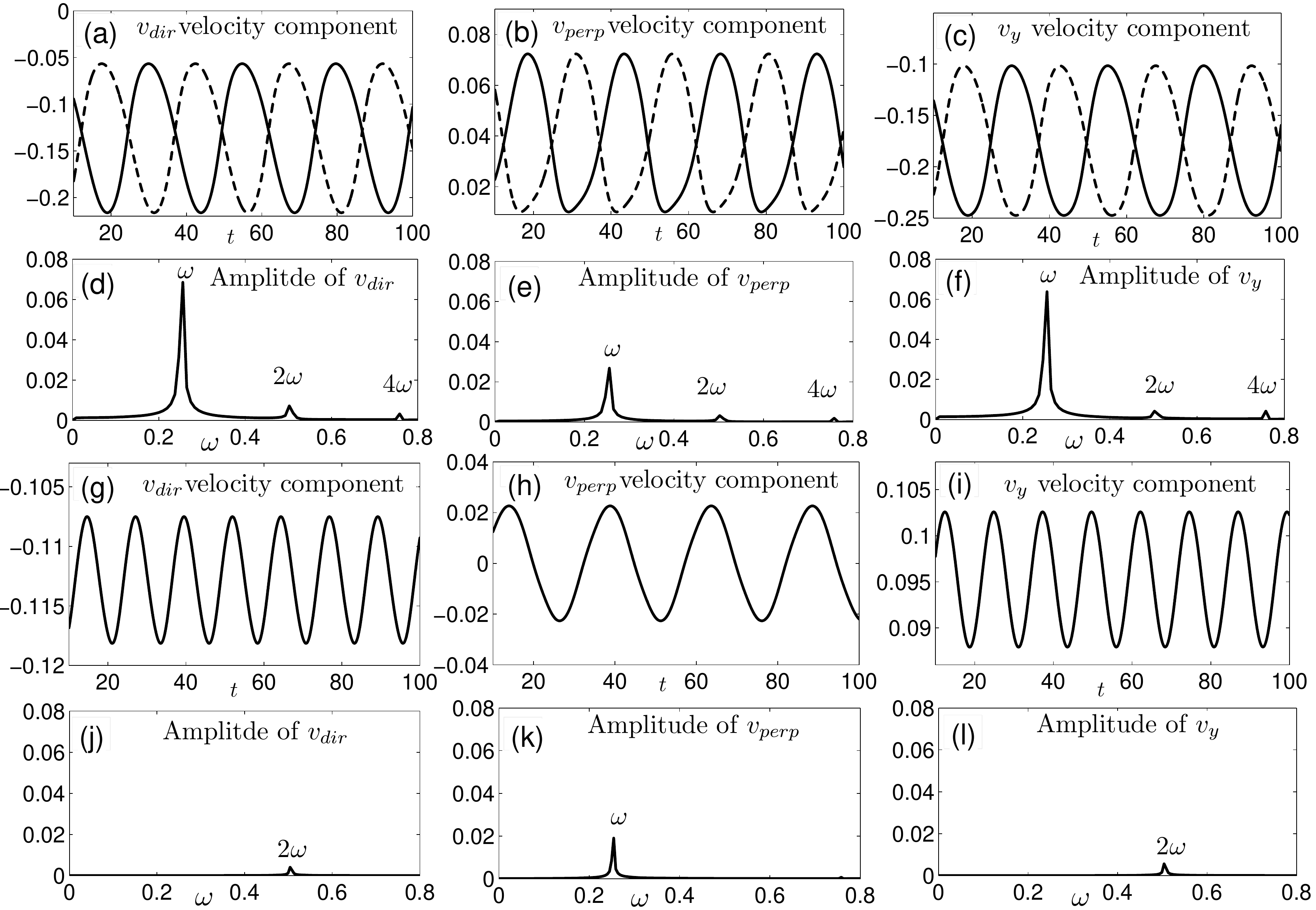}
\end{center}
\caption{Time evaluations of $v_{dir}$, $v_{perp}$, $v_{y}$ velocity components and the corresponding frequency spectra, $Re=2335$: (a)-(f) solid line - control point A(0.66, 0.42, 0.56), dashed line - control point B(0.56 0.42,0.66); (g)-(l) control point C(0.1 0.14 0.1) }
\label{fig:evaluations}
\end{figure}
 We now focus on an instantaneous relationship between the flow fields from both sides of the cavity main diagonal plane. It can be clearly seen that signals for all three velocity components monitored in points A and B (see Figs. \ref{fig:evaluations}(a-c) are out of phase up to a small offset. This indicates an opposite flow direction in these points at any given time. The observed offset is a consequence of a symmetry break resulting from an instantaneous deformation of the interface surface separating the two cavity parts. As expected, all three velocity components are characterized by the same spectra consisting of a leading harmonic and its multipliers (see Figs. \ref{fig:evaluations}(d-f)). The $v_{dir}$ and $v_{y}$ velocity components have close values of oscillating amplitude, which are about twice as large as that of $v_{perp}$. Close to the cavity main diagonal plane the offset between time signals monitored in any pair of  symmetry reflection points approach zero and both time evaluations become out of phase. Therefore only a doubled harmonic is observed for the $v_{dir}$ and $v_{y}$ signals monitored at point C, while the basic harmonic is canceled out. At the same time the later persists in the $v_{perp}$ velocity component characterizing oscillations of the interface surface (see Figs. \ref{fig:evaluations}(g-l)).

\subsection{Oscillation amplitude analysis}
The fact that the developed supercritical flow ($Re=2335$) is determined by only a single oscillating mode allowed us to calculate the flow oscillation amplitudes for all velocity components. This was done by computing  a deviation between the maximum and the base flow values of a given velocity component attained at each grid point and averaged over several oscillation periods. Next we plotted isosurfaces confining the regions where the oscillation amplitude values are no less than 25\% of the maximal amplitude of the corresponding velocity component, as shown in Figs. \ref{fig:deviations} a-c. The contours of maximal amplitudes in spanwise and main diagonal cross sections are shown in Figs. \ref{fig:deviations} d-f and Figs.\ref{fig:deviations} g-i, respectively. The procedure comprises a convenient way to determine of the most energetic flow regions (see e.g. \cite{Theofilis2005JFM}, and \cite{feldman2010physfluid}). It should be noted that because of the subcritical character of the  bifurcation the calculated oscillation amplitudes can not be mathematically related to the absolute values of the flow eigenvectors. Nevertheless, given a good agreement between the previous DNS study of \cite{feldman2010physfluid} (performed for a ``classical'' lid-driven cavity) and the corresponding linear stability analysis of \cite{gianettiatti}, a striking resemblance between both is substantiated.
\begin{figure}
\centering
    \subfigure[]
    {
        \includegraphics[width=0.29\textwidth,clip=]{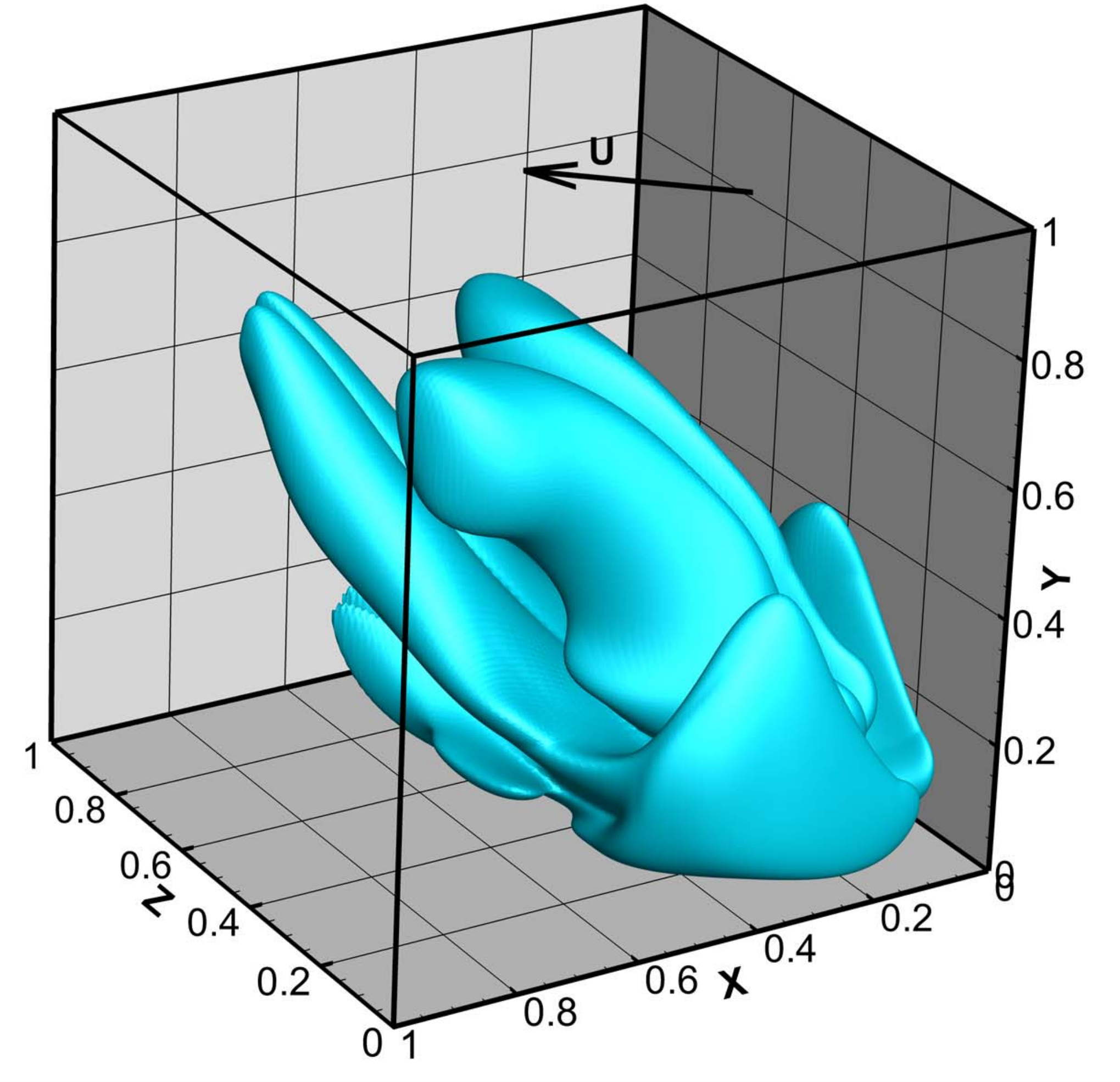}
    }
    \subfigure[]
    {
        \includegraphics[width=0.29\textwidth,clip=]{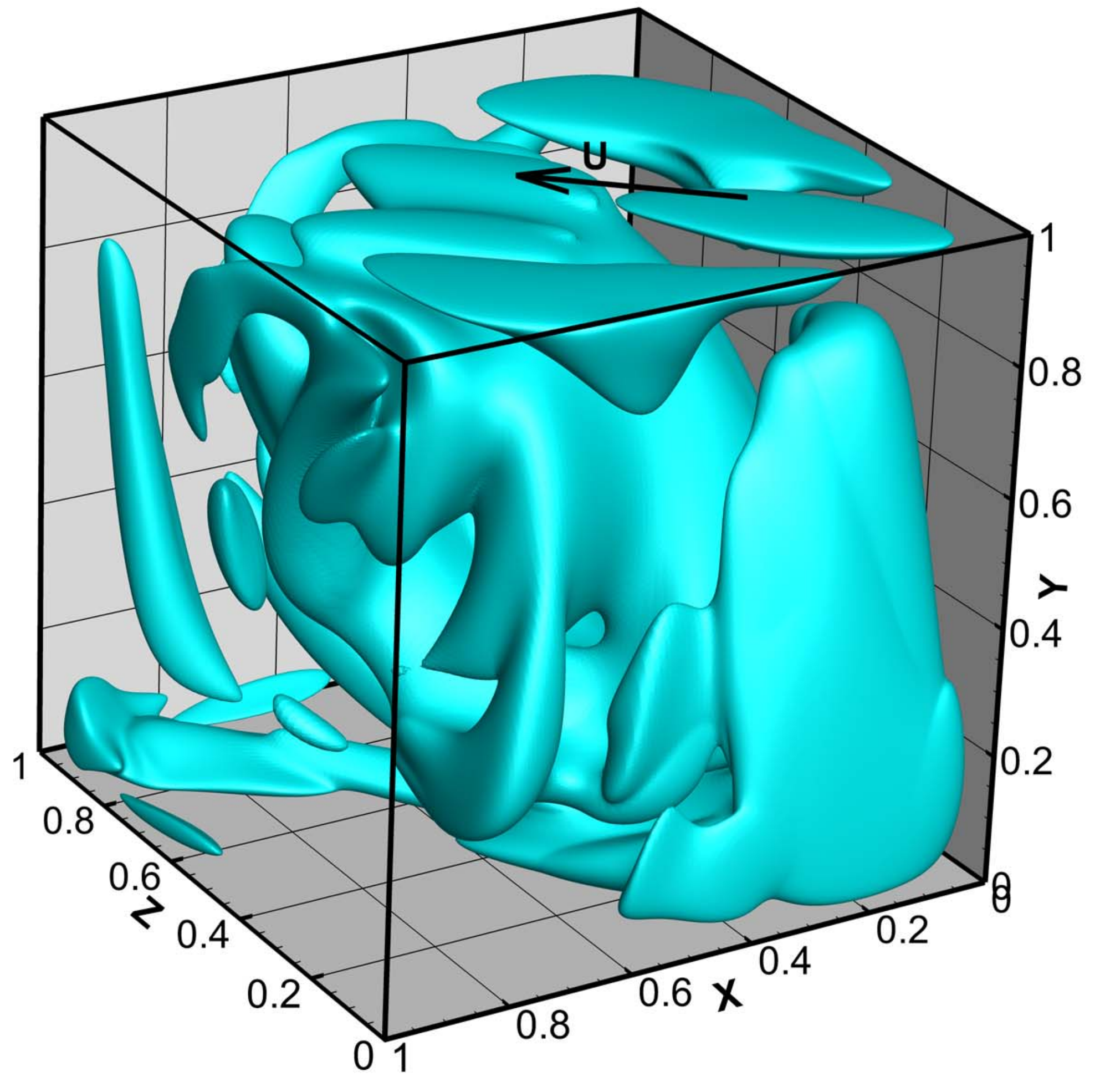}
    }
    \subfigure[]
    {
        \includegraphics[width=0.29\textwidth,clip=]{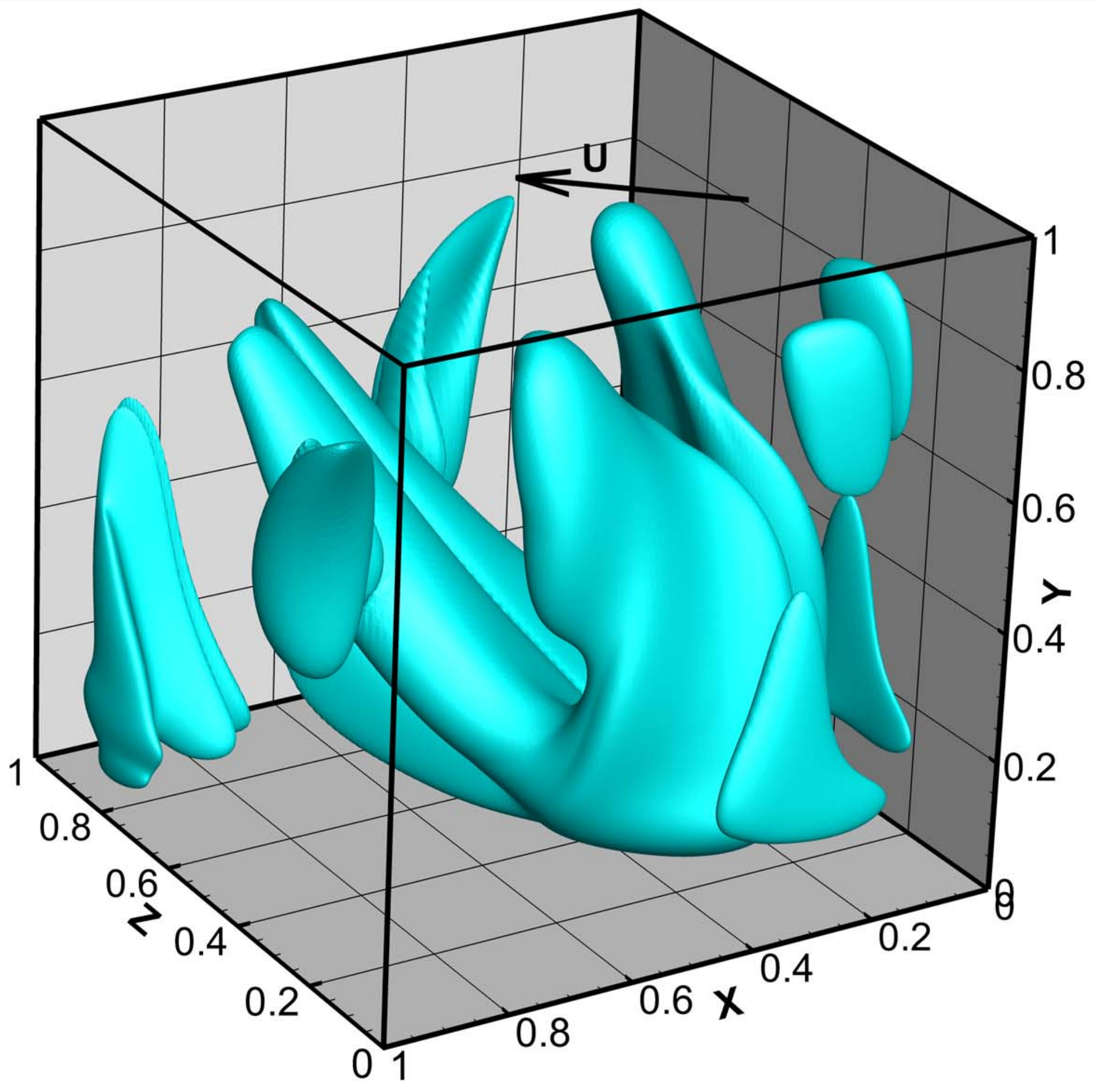}
    }

    \subfigure[]
    {
        \includegraphics[width=0.29\textwidth,clip=]{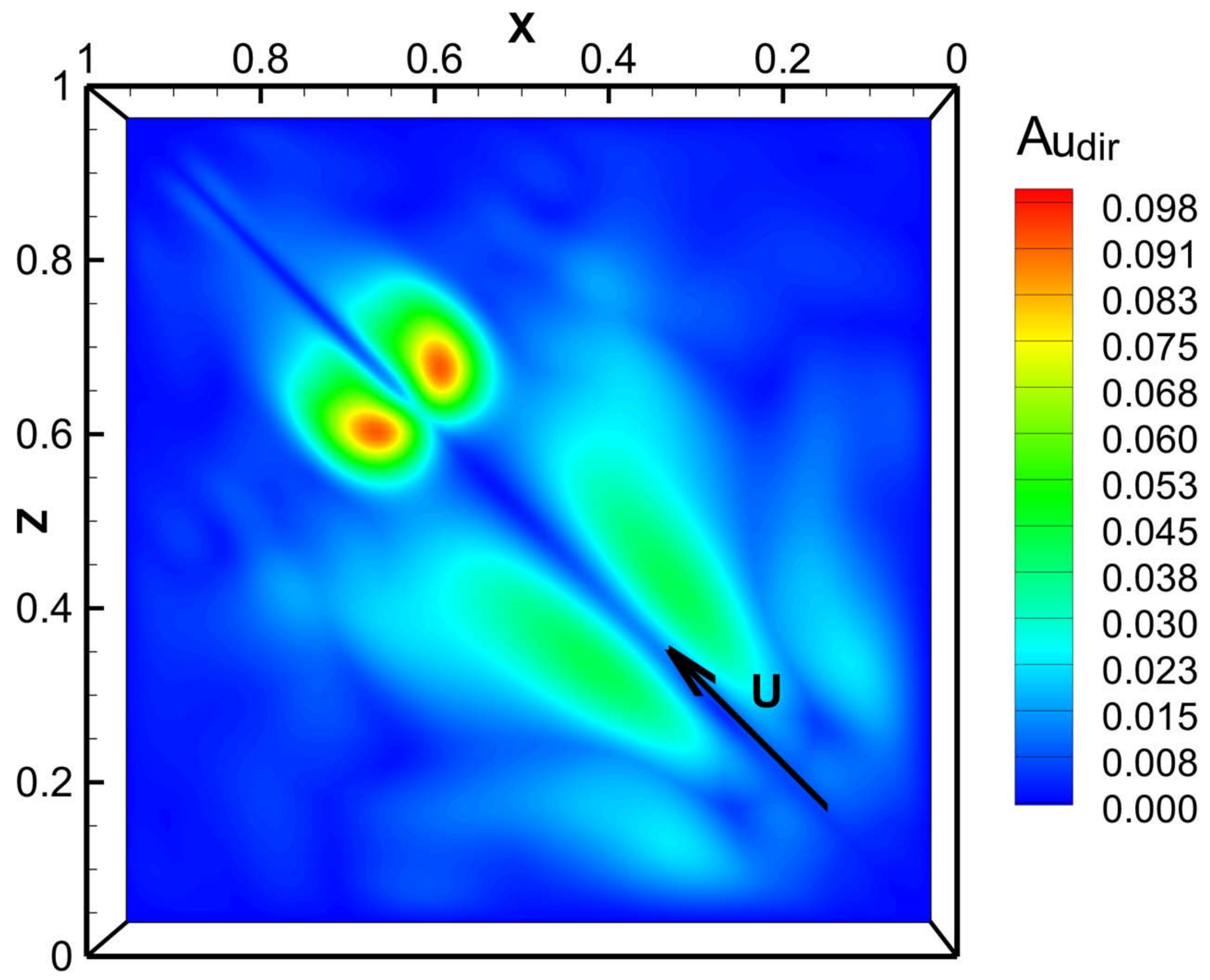}
    }
    \subfigure[]
    {
        \includegraphics[width=0.29\textwidth,clip=]{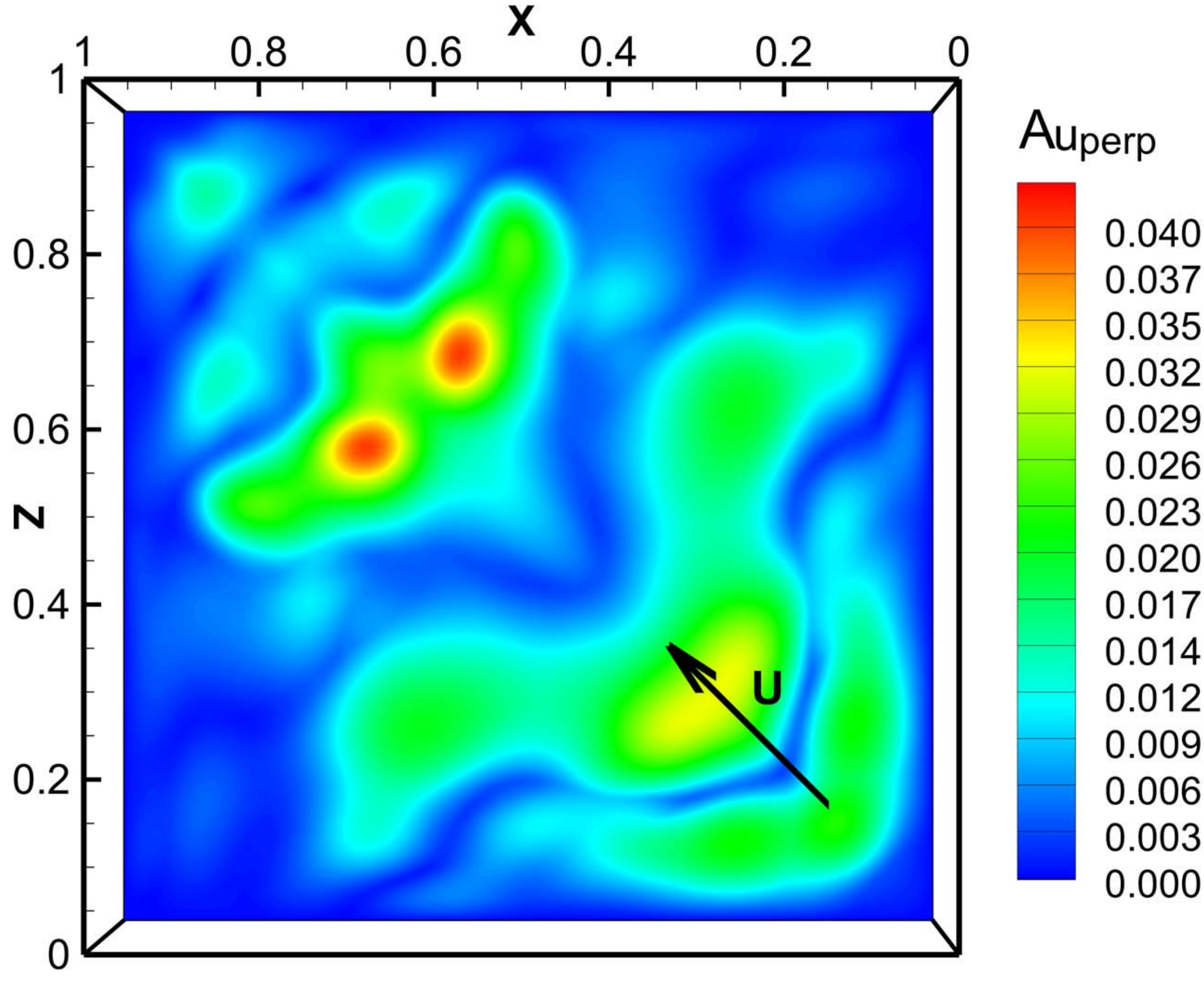}
    }
    \subfigure[]
    {
        \includegraphics[width=0.29\textwidth,clip=]{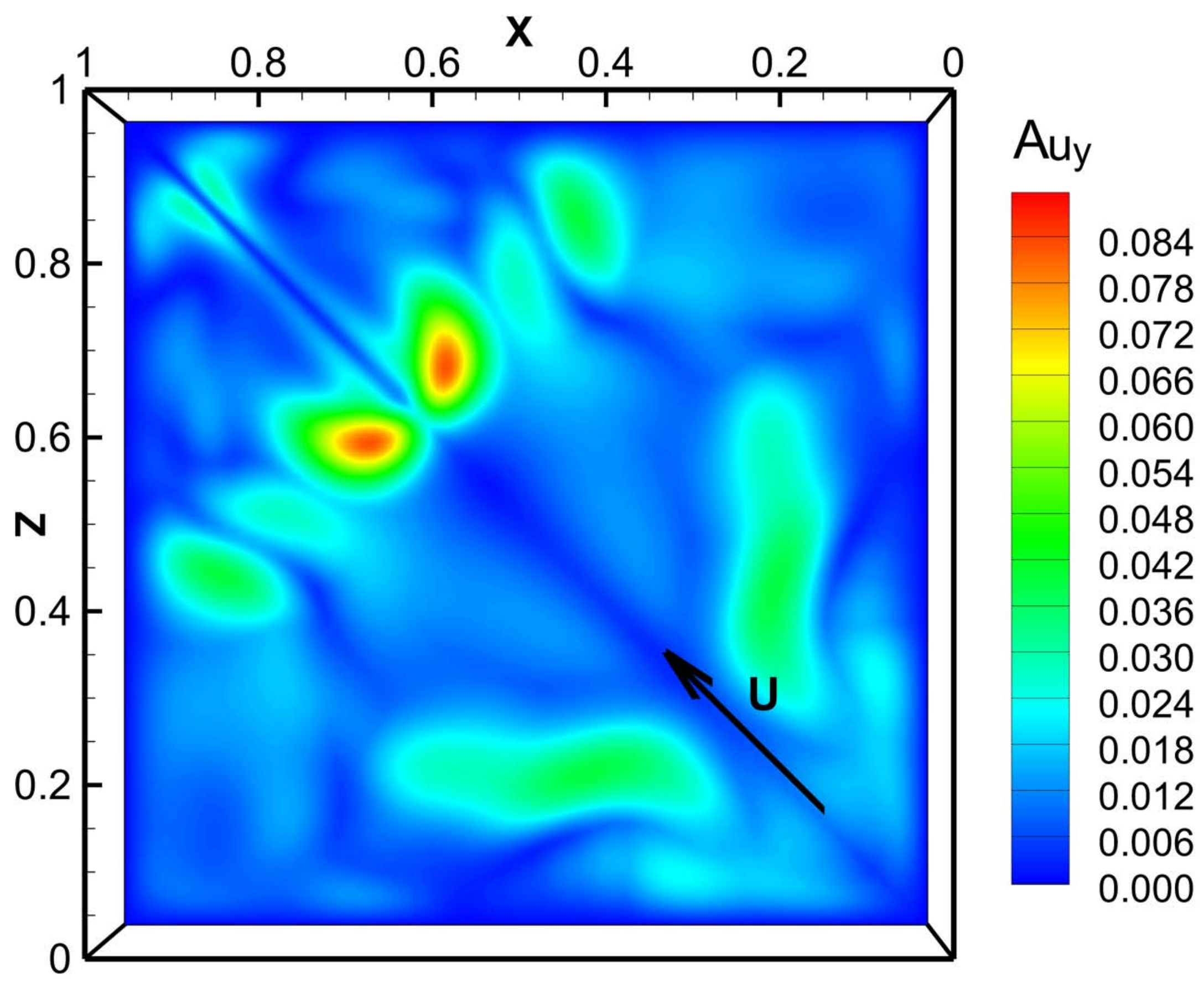}
    }

    \subfigure[]
    {
        \includegraphics[width=0.29\textwidth,clip=]{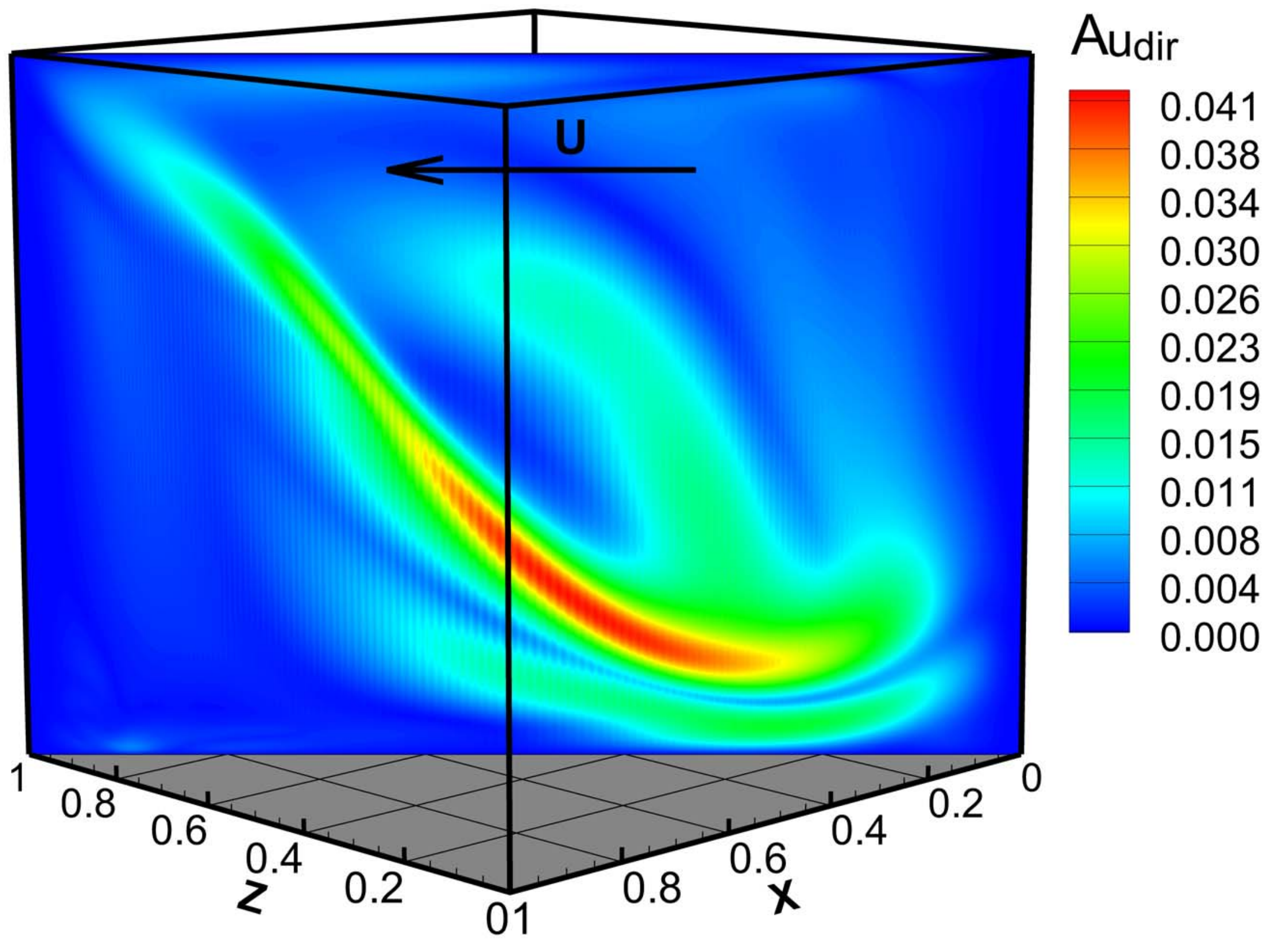}
    }
    \subfigure[]
    {
        \includegraphics[width=0.29\textwidth,clip=]{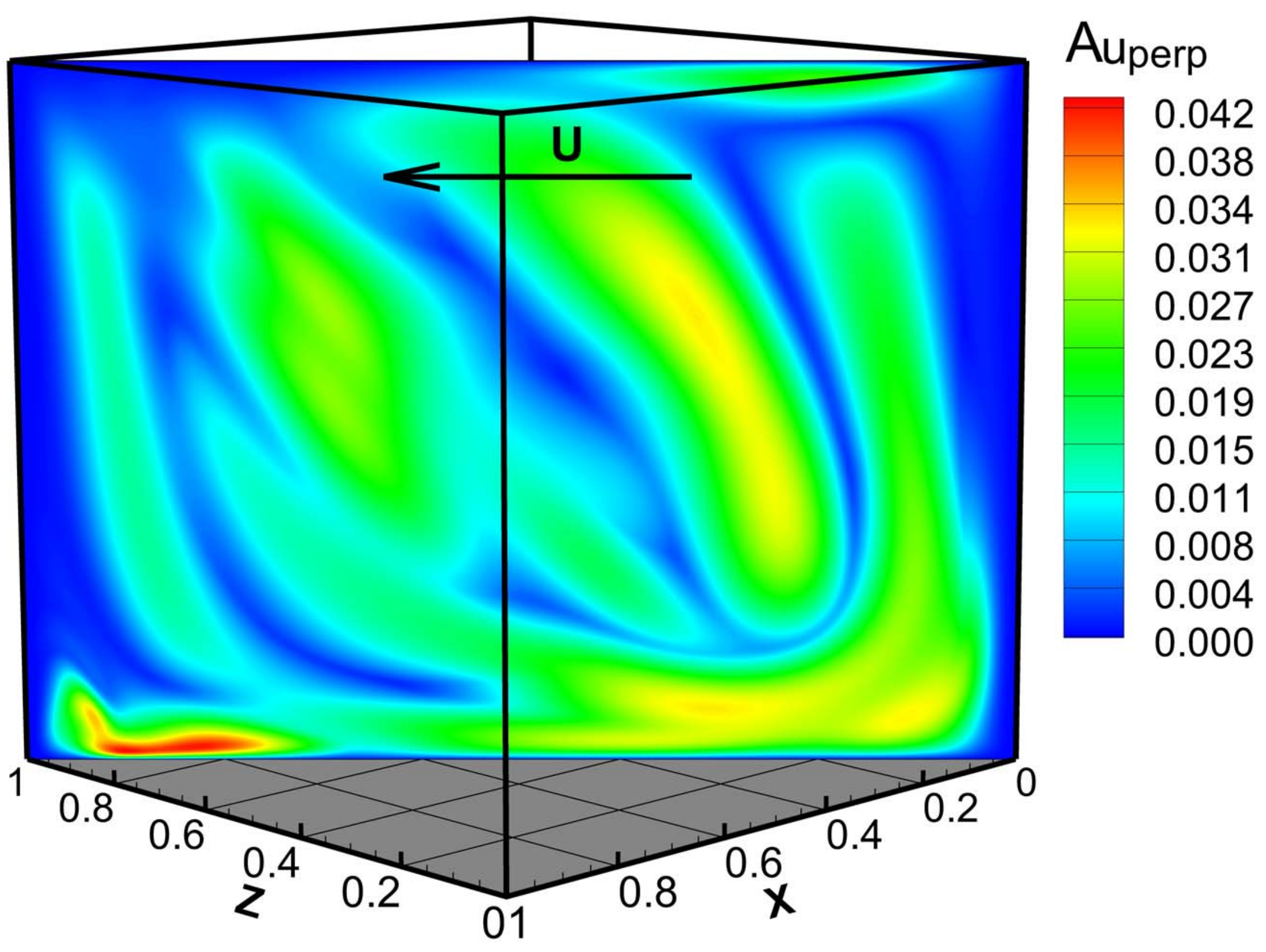}
    }
    \subfigure[]
    {
        \includegraphics[width=0.29\textwidth,clip=]{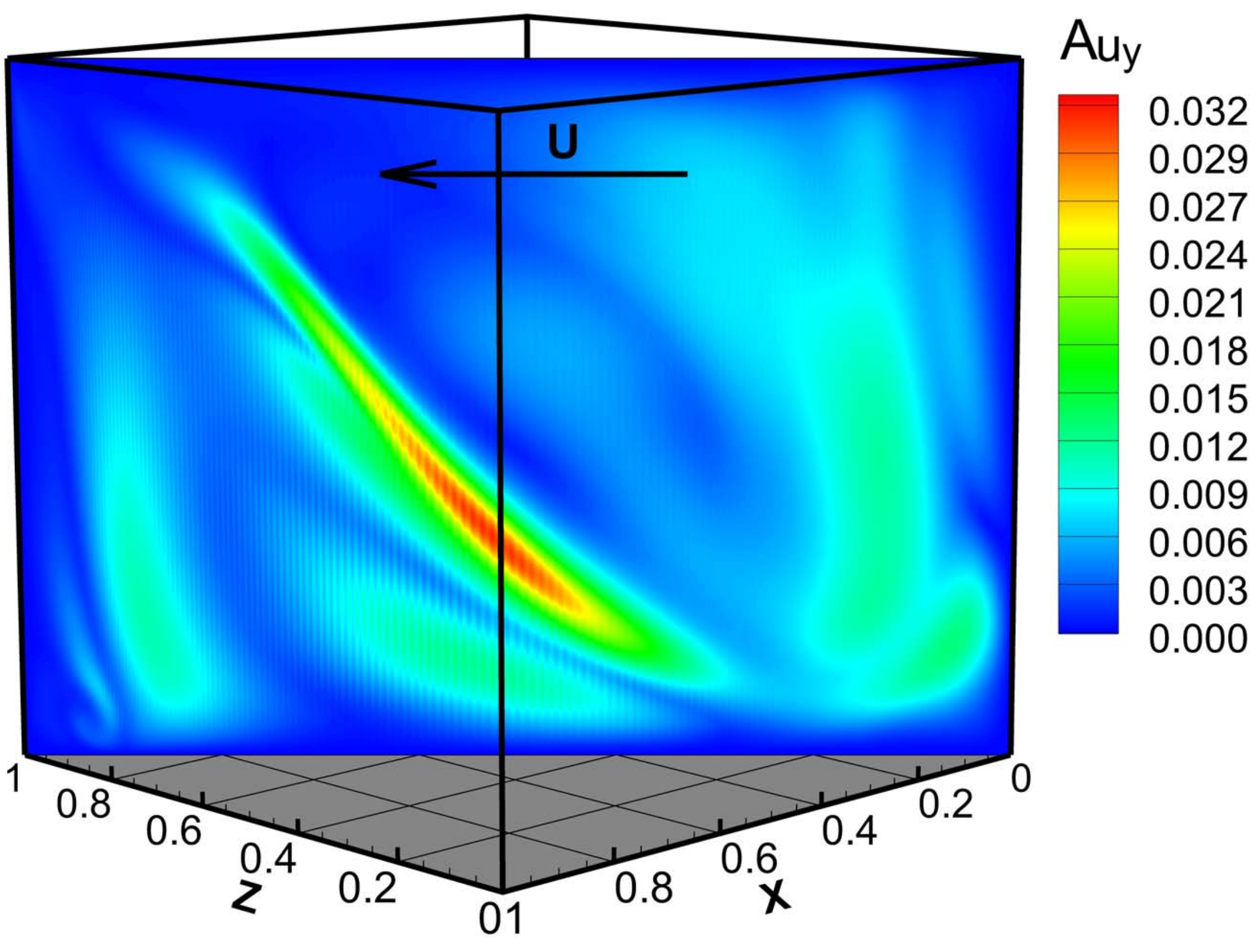}
    }
\caption{Spatial distribution of oscillations amplitudes for the $u_{dir}$,$u_{perp}$,$u_{y}$ velocity components, lid moves as indicated by an arrow: (a),(b),(c) 3-D contours confining the areas with $A\geq0.25A_{max}$; (d),(e),(f) values obtained in the horizontal midplane; (h),(i),(j) values obtained in the main diagonal cross-section.}
\label{fig:deviations}
\end{figure}

As can be easily recognized, the spatial pattern of the amplitude values is reflection symmetric with respect to the cavity main diagonal plane. Maximum values of all oscillation amplitudes are side biased from the main diagonal surface. At the same time, one can observe substantial qualitative differences between the spatial distribution of directional and vertical amplitudes $A_{u_{dir}}$ and $A_{u_{y}}$  compared to that of the perpendicular amplitude $A_{u_{perp}}$.  First, $A_{u_{dir}}$ and $A_{u_{y}}$ are compactly grouped around the main diagonal plane, while $A_{u_{perp}}$ distribution is quite dispersed and occupies the most of cavity volume (see Figs. \ref{fig:deviations} a-c). Second, the  maxima of $A_{u_{dir}}$ and $A_{u_{y}}$ are about twice as higher as that of $A_{u_{perp}}$ as can be seen from Figs. \ref{fig:deviations} d-f. Third, maximal values of $A_{u_{dir}}$ and $A_{u_{y}}$ oscillation amplitudes observed on a spanwise cross-section plane are more than twice as high as those found on the main diagonal plane, while $A_{u_{perp}}$ has about the same values on both planes. The observed differences indicate a dominant role of $u_{dir}$ and $u_{y}$ oscillations in the dynamics of the developed supercritical flow.

\subsection{Instability characteristics}
Investigation of instabilities of internal vortices in driven cavities occupies a prominent place in state of the art experimental and CFD research. The complexity of the vortex interactions effected by the presence of non-slip boundaries gives rise to an impressive variety of instabilities and bifurcation scenarios \cite{kuhlmann2005MM}. In the following, we focus our attention on oscillatory instability, observed in the present slightly supercritical flow.

Figure \ref{fig:vorticity}(a) presents a \textit{y} (vertical) component of vorticity distribution of the base flow taken at the cavity midspan cross section and computed as a curl of velocity field averaged over a whole number of periods. The contours are superimposed by the representative base flow pathlines projected on to the midspan plane. The pattern thus obtained can be interpreted as a flow \textit{topographic map}, in which the closed contours determine the size and position of the corresponding vortex cores. Four different vortex groups (marked by numbers 1-4) can be recognized bearing a striking resemblance to the steady state flow pattern  reported for $Re=2000$ in \cite{povitsky2005}. It should be noted that only the counter-rotating vortex pair marked by the number 2 has a clearly distinguished core of elliptic shape, whereas all other vortices exhibit the flow converging to their centers. The averaged vorticity field is antisymmetric and strained close to the diagonal interface plane where it attains maximum absolute values, similar to that, determined in \cite{leweke1998JFM} with respect to stationary elliptic instability of the vortex pair in the open flow.

As already mentioned, given that the flow is determined by only a single oscillating mode with moderate amplitudes, the spatial distribution of eigenvector magnitudes of any flow field can be approximated by the corresponding oscillation amplitude distribution (see e.g., \cite{feldman2010physfluid}). Fig. \ref{fig:vorticity}(b) presents the reflection symmetric distribution of the oscillation amplitude of \textit{y} vorticity component taken at the midspan cross section. It is remarkable that the diameter of an area confined by a white dashed line corresponding to approximately zero value of the oscillating amplitude (and therefore to the zero of $\omega_y$ perturbation) is about twice that of the vortex core. The observed relationship is not just a coincidence and was verified in a number of additional span cross sections where the existence of the vortex pair core can still be recognized (in the range of  $0.45\leq y \leq 0.6$). The later comprises an important invariant relationship with respect to the perturbation field of the vortex pair and repeats the observation made for elliptic instabilyty \cite{leweke1998JFM}. At the same time, the characteristic two-lobe structure of the perturbation for each vortex theoretically predicted by \cite{waleffe1990physfluid} and verified by \cite{leweke1998JFM} can not be reproduced by oscillation amplitudes distribution since the later is only related to the absolute value of perturbation.

\begin{figure}
\centering
    \subfigure[]
    {
        \includegraphics[width=0.43\textwidth,clip=]{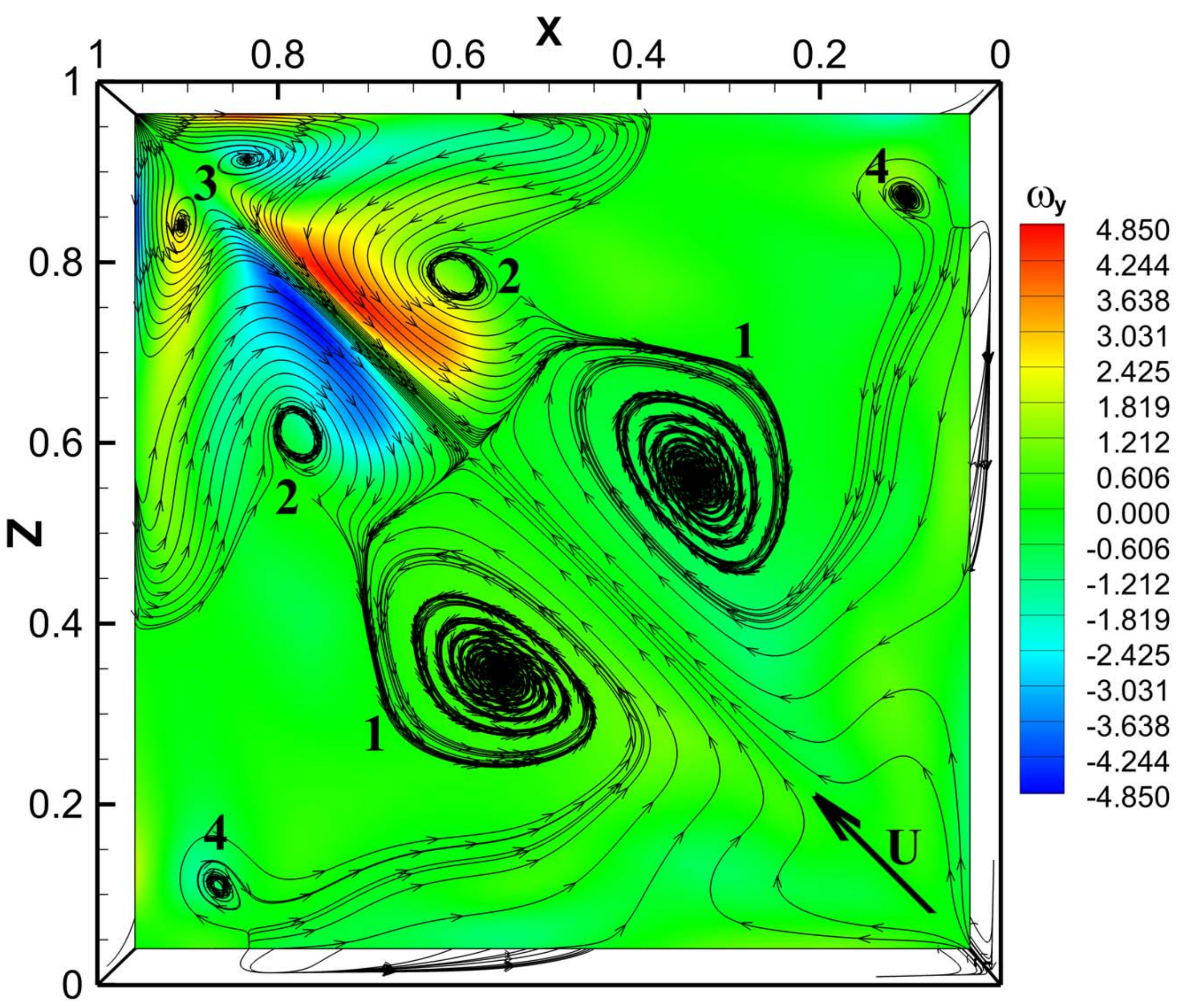}
            }
    \subfigure[]
    {
        \includegraphics[width=0.43\textwidth,clip=]{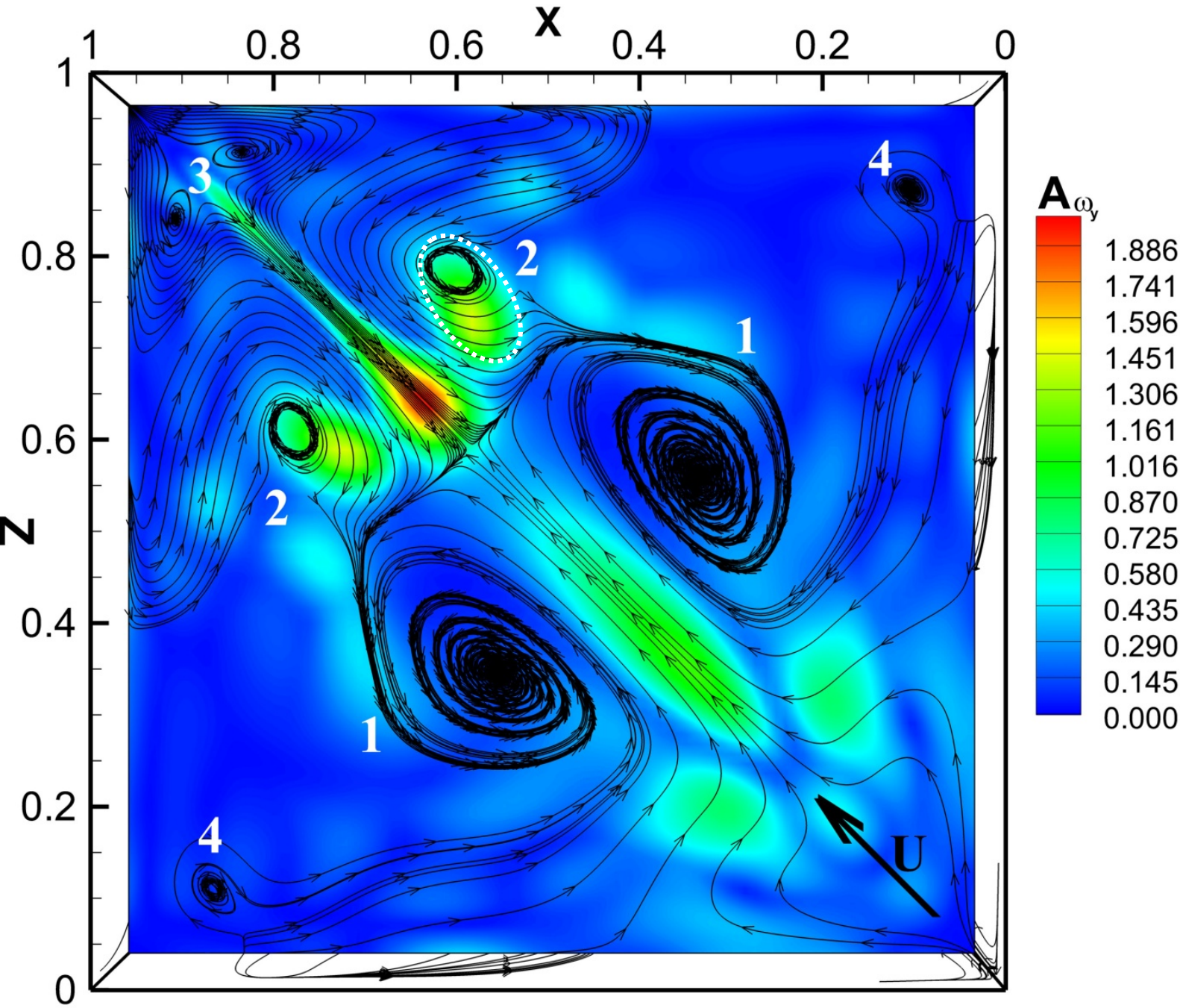}
    }
    \caption{Base flow path lines projected on a mid cross section, superimposed with: (a) base (averaged) vorticity field; (b) oscillation amplitudes of $y$ (vertical) component of the vorticity field. Lid moves as indicated by an arrow.}
\label{fig:vorticity}
\end{figure}

It should be noted that in spite of the given qualitative arguments in favor of elliptic character of the observed instability its actual nature is still to be determined. Generally, elliptic instability can exhibit oscillatory modes (see e.g. \cite{Billant1999physfluid, Roy2008physfluid, Donnadieu2009physfluid}) but their  structure is more complex than that observed for the stationary elliptic instability by \cite{leweke1998JFM}. The flow dynamics is also complicated by the presence of boundaries which effects the mechanism of mutual vortices interaction. Looking at Fig.\ref{fig:vorticity}(b) one can recognize the existence of an oscillation amplitude peak of $\omega_{y}$  located on the interface between the two vortices which is apparently the result of local centrifugal effects of the flow. In this region, the streams symmetrically moving away from the cavity walls meet the oppositely directed stream driven by the cavity lid (see Fig. \ref{fig:vorticity}(b) regions 1, and 2). As a result, both counter-flows are decelerating while turning downward into the cavity and then each separately going to its cavity part. At the same time inertia interferes with the flow rotation originating instability characterized by increased velocity and vorticity oscillations. The quantitative characteristics of the observed instability like $Q$ criterion, the strain, the ellipticity and the growth rate can shed light on the actual nature of the observed instability and will be in the focus of our future study.

\section{Conclusions}
An oscillatory slightly supercritical fully 3D flow in a diagonally lid-driven cavity was thoroughly investigated. A series of time dependent DNS computations was used to determine critical $Re_{cr}$ and $\omega_{cr}$ values for transition from steady to oscillatory flow. The calculations were performed on two successive stretched grids of $100^3$ and $200^3$ finite volumes and Richardson extrapolation was used to approximate the results to the zero-grid-size limit. It was found that the transition to unsteadiness takes place via subcritical Hopf bifurcation at $Re_{cr}=2320$ and $\omega_{cr}=0.249$ and is characterized by a symmetry break of the flow.

The slightly supercritical flow is characterized by reflection symmetric fields of velocity base flow and the oscillation amplitudes of both velocity and vorticity relatively to the main diagonal interface surface. In contrast the vorticity base flow is antisymmetric. The velocity fields from both sides of the main diagonal plane oscillate out of phase up to a small offset. The later is a consequence of a symmetry break resulting from an oscillation of the interface plane. As the distance to the interface surface is decreased, the offset value approaches to zero.

The mechanism of the subcritical flow instability was studied. It was found that the instability is of oscillatory type and evolves on two vortices in a coupled manner. The instability observed in a confined cubic enclosure enjoys the same qualitative characteristics as that experimentally observed by \cite{leweke1998JFM} for the open flow vortex pair, namely elliptic shape of the streamlines in the core of each vortex, internal deformation of the vortex cores, a distinct phase relationship between the perturbations on each vortex. Also characteristic is the approximate 2:1 ratio of the size of vortex perturbation to the size of the vortex core, recognized by \cite{leweke1998JFM} as an important invariant value. At the same time, more comprehensive quantitative analysis to be performed in the future to determine the actual nature of the observed instability. The present study represents a signifficant milestone in the global stability analysis of fully confined recirculating flows and is of great importance for verification of future experimental and numerical results.
\\

\label{}





\bibliographystyle{model1a-num-names}
\bibliography{jfm-instructions}

\begin{thebibliography}{30}
\expandafter\ifx\csname natexlab\endcsname\relax\def\natexlab#1{#1}\fi
\providecommand{\bibinfo}[2]{#2}
\ifx\xfnm\relax \def\xfnm[#1]{\unskip,\space#1}\fi
\bibitem[{Shankar and Deshpande(2000)}]{shankar2000anrev}
\bibinfo{author}{P.~N. Shankar}, \bibinfo{author}{M.~Deshpande},
  \bibinfo{journal}{Annual Review of Fluid Dynamics} \bibinfo{volume}{32}
  (\bibinfo{year}{2000}) \bibinfo{pages}{93--136}.
\bibitem[{Batchelor(1956)}]{batchelor1956jfm}
\bibinfo{author}{G.~K. Batchelor}, \bibinfo{journal}{J. Fluid Mech.}
  \bibinfo{volume}{1} (\bibinfo{year}{1956}) \bibinfo{pages}{177--190}.
\bibitem[{Moffatt(1963)}]{moffatt1963jfm}
\bibinfo{author}{H.~K. Moffatt}, \bibinfo{journal}{J. Fluid Mech.}
  \bibinfo{volume}{18} (\bibinfo{year}{1963}) \bibinfo{pages}{1--18}.
\bibitem[{Kawaguti(1961)}]{kawaguti1961jpsj}
\bibinfo{author}{M.~Kawaguti}, \bibinfo{journal}{J. Phys. Soc. Japan}
  \bibinfo{volume}{16} (\bibinfo{year}{1961}) \bibinfo{pages}{2307--2315}.
\bibitem[{Simuni(1965)}]{simuni1965jamtp}
\bibinfo{author}{L.~M. Simuni}, \bibinfo{journal}{J. Appl. Mech. Tech. Phys.}
  \bibinfo{volume}{6} (\bibinfo{year}{1965}) \bibinfo{pages}{106--108}.
\bibitem[{Feldman and Gelfgat(2010)}]{feldman2010physfluid}
\bibinfo{author}{Y.~Feldman}, \bibinfo{author}{A.~Y. Gelfgat},
  \bibinfo{journal}{Physics of Fluids} \bibinfo{volume}{22}
  (\bibinfo{year}{2010}) \bibinfo{pages}{093602}.
\bibitem[{Povitsky(2005)}]{povitsky2005}
\bibinfo{author}{A.~Povitsky}, \bibinfo{journal}{Nonlinear Analysis}
  \bibinfo{volume}{63} (\bibinfo{year}{2005}) \bibinfo{pages}{e1573--e1584}.
\bibitem[{Feldman and Gelfgat(2010)}]{feldman2010computfluid}
\bibinfo{author}{Y.~Feldman}, \bibinfo{author}{A.~Y. Gelfgat},
  \bibinfo{journal}{Computers\&Fluids} \bibinfo{volume}{11}
  (\bibinfo{year}{2010}) \bibinfo{pages}{218--223}.
\bibitem[{Yu et~al.(2003)Yu, Mei, Luo, and Shyy}]{dazhi2003}
\bibinfo{author}{D.~Yu}, \bibinfo{author}{R.~Mei}, \bibinfo{author}{L.-S. Luo},
  \bibinfo{author}{W.~Shyy}, \bibinfo{journal}{Progress in Aerosp. Sc.}
  \bibinfo{volume}{39} (\bibinfo{year}{2003}) \bibinfo{pages}{329--367}.
\bibitem[{D'Humi{\`{e}}res et~al.(2012)D'Humi{\`{e}}res, Ginzburg, Krafczyk,
  Lallemand, and Luo}]{dhumi2011PhilTrans}
\bibinfo{author}{D.~D'Humi{\`{e}}res}, \bibinfo{author}{I.~Ginzburg},
  \bibinfo{author}{M.~Krafczyk}, \bibinfo{author}{P.~Lallemand},
  \bibinfo{author}{L.-S. Luo}, \bibinfo{journal}{Phil. Trans. R. Soc. Lond. A}
  \bibinfo{volume}{360} (\bibinfo{year}{2012}) \bibinfo{pages}{437--451}.
\bibitem[{Freitas et~al.(2011)Freitas, Henze, Meinke, and
  Schr{\"{o}}der}]{freitas2011CompFluids}
\bibinfo{author}{R.~K. Freitas}, \bibinfo{author}{A.~Henze},
  \bibinfo{author}{M.~Meinke}, \bibinfo{author}{W.~Schr{\"{o}}der},
  \bibinfo{journal}{Computers\&Fluids} \bibinfo{volume}{47}
  (\bibinfo{year}{2011}) \bibinfo{pages}{115--121}.
\bibitem[{Asinari et~al.(2012)Asinari, Ohwada, Chiavazzo, and
  Di~Rienzo}]{asinari2012jcp}
\bibinfo{author}{P.~Asinari}, \bibinfo{author}{T.~Ohwada},
  \bibinfo{author}{E.~Chiavazzo}, \bibinfo{author}{A.~F. Di~Rienzo},
  \bibinfo{journal}{Journal of Computational Physics} \bibinfo{volume}{231}
  (\bibinfo{year}{2012}) \bibinfo{pages}{5109--5143}.
\bibitem[{Ramanan and Homsy(1994)}]{ramanan1994physfl}
\bibinfo{author}{N.~Ramanan}, \bibinfo{author}{G.~M. Homsy},
  \bibinfo{journal}{Physics of Fluids} \bibinfo{volume}{6}
  (\bibinfo{year}{1994}) \bibinfo{pages}{2690--2701}.
\bibitem[{Theofilis(2000)}]{theofilis2000AIAA}
\bibinfo{author}{V.~Theofilis}, \bibinfo{journal}{AIAA paper}
  (\bibinfo{year}{2000}) \bibinfo{pages}{2000--1965}.
\bibitem[{Ding and Kawahara(1998)}]{ding1998JCP}
\bibinfo{author}{Y.~Ding}, \bibinfo{author}{M.~Kawahara}, \bibinfo{journal}{J.
  Comput. Phys.} \bibinfo{volume}{139} (\bibinfo{year}{1998})
  \bibinfo{pages}{243--273}.
\bibitem[{Albensoeder and Kuhlmann(2001)}]{albensoeder2001physfluid}
\bibinfo{author}{S.~Albensoeder}, \bibinfo{author}{H.~Kuhlmann},
  \bibinfo{journal}{Physics of Fluids} \bibinfo{volume}{13}
  (\bibinfo{year}{2001}) \bibinfo{pages}{121--136}.
\bibitem[{Giannetti et~al.(2009)Giannetti, Luchini, and Marino}]{gianettiatti}
\bibinfo{author}{F.~Giannetti}, \bibinfo{author}{P.~Luchini},
  \bibinfo{author}{L.~Marino}, in: \bibinfo{booktitle}{{ATTI 19th Congr. AIMETA
  Mecc. Teor. Appl.}}, \bibinfo{address}{Ancona, Italy}.
\bibitem[{Liberzon et~al.(2011)Liberzon, Feldman, and
  Gelfgat}]{liberzon2011physfluid}
\bibinfo{author}{A.~Liberzon}, \bibinfo{author}{Y.~Feldman},
  \bibinfo{author}{A.~Y. Gelfgat}, \bibinfo{journal}{Physics of Fluids}
  \bibinfo{volume}{23} (\bibinfo{year}{2011}) \bibinfo{pages}{084106}.
\bibitem[{Weller et~al.(1998)Weller, Tabor, Jasak, and Fureby}]{weller1998}
\bibinfo{author}{H.~Weller}, \bibinfo{author}{G.~Tabor},
  \bibinfo{author}{H.~Jasak}, \bibinfo{author}{C.~Fureby},
  \bibinfo{journal}{Computers in Physics} \bibinfo{volume}{12}
  (\bibinfo{year}{1998}) \bibinfo{pages}{620--631}.
\bibitem[{Ferziger and Peri{\'c}(2002)}]{ferzirger2002}
\bibinfo{author}{J.~H. Ferziger}, \bibinfo{author}{M.~Peri{\'c}},
  \bibinfo{title}{Computational methods for fluid dynamics},
  \bibinfo{publisher}{Springer}, \bibinfo{year}{2002}.
\bibitem[{Drazin and Reid(2004)}]{drazin2004}
\bibinfo{author}{P.~Drazin}, \bibinfo{author}{W.~Reid},
  \bibinfo{title}{Hydrodynamic stability}, \bibinfo{publisher}{Cambridge
  Mathematical Library}, \bibinfo{year}{2004}.
\bibitem[{Hassard et~al.(1981)Hassard, Kazarinoff, and Wan}]{hassard1981}
\bibinfo{author}{B.~Hassard}, \bibinfo{author}{N.~Kazarinoff},
  \bibinfo{author}{Y.-H. Wan}, \bibinfo{title}{Theory and applications of Hopf
  bifurcation}, \bibinfo{publisher}{Mathematics Society Lecture Note Series
  Vol. 41 London}, \bibinfo{year}{1981}.
\bibitem[{Gelfgat(2007)}]{gelfgat2007Int_j}
\bibinfo{author}{A.~Y. Gelfgat}, \bibinfo{journal}{Int. J. Numer Methods
  Fluids} \bibinfo{volume}{53} (\bibinfo{year}{2007})
  \bibinfo{pages}{485--506}.
\bibitem[{Theofilis et~al.(2005)Theofilis, Duck, and Owen}]{Theofilis2005JFM}
\bibinfo{author}{V.~Theofilis}, \bibinfo{author}{P.~W. Duck},
  \bibinfo{author}{J.~Owen}, \bibinfo{journal}{Journal of Fluid Mechanics}
  \bibinfo{volume}{505} (\bibinfo{year}{2005}) \bibinfo{pages}{249--286}.
\bibitem[{Kuhlmann and Albensoeder(2005)}]{kuhlmann2005MM}
\bibinfo{author}{H.~Kuhlmann}, \bibinfo{author}{S.~Albensoeder},
  \bibinfo{journal}{Z Angew Math Mech} \bibinfo{volume}{85}
  (\bibinfo{year}{2005}) \bibinfo{pages}{387--399}.
\bibitem[{Leweke and Williamson(1998)}]{leweke1998JFM}
\bibinfo{author}{T.~Leweke}, \bibinfo{author}{C.~H.~K. Williamson},
  \bibinfo{journal}{Journal of Fluid Mechanics} \bibinfo{volume}{360}
  (\bibinfo{year}{1998}) \bibinfo{pages}{85--119}.
\bibitem[{Waleffe(1994)}]{waleffe1990physfluid}
\bibinfo{author}{J.~Waleffe}, \bibinfo{journal}{Physics of Fluids A}
  \bibinfo{volume}{2} (\bibinfo{year}{1994}) \bibinfo{pages}{76--80}.
\bibitem[{Billant et~al.(1999)Billant, Brancher, and
  Chomaz}]{Billant1999physfluid}
\bibinfo{author}{P.~Billant}, \bibinfo{author}{P.~Brancher},
  \bibinfo{author}{J.-M. Chomaz}, \bibinfo{journal}{Physics of Fluids}
  \bibinfo{volume}{11} (\bibinfo{year}{1999}) \bibinfo{pages}{2069--2077}.
\bibitem[{Roy et~al.(2008)Roy, Schaeffer, Le~Diz{\`e}s, and
  Thompson}]{Roy2008physfluid}
\bibinfo{author}{C.~Roy}, \bibinfo{author}{N.~Schaeffer},
  \bibinfo{author}{S.~Le~Diz{\`e}s}, \bibinfo{author}{M.~Thompson},
  \bibinfo{journal}{Physics of Fluids} \bibinfo{volume}{20}
  (\bibinfo{year}{2008}) \bibinfo{pages}{094101}.
\bibitem[{Donnadieu et~al.(2009)Donnadieu, Ortiz, Chomaz, and
  P.}]{Donnadieu2009physfluid}
\bibinfo{author}{C.~Donnadieu}, \bibinfo{author}{S.~Ortiz},
  \bibinfo{author}{J.-M. Chomaz}, \bibinfo{author}{B.~P.},
  \bibinfo{journal}{Physics of Fluids} \bibinfo{volume}{21}
  (\bibinfo{year}{2009}) \bibinfo{pages}{094102}.

\end{thebibliography}







\end{document}